\documentclass[aps,prl,twocolumn,showpacs]{revtex4}

\usepackage{amsmath,amssymb,units,graphicx}

\graphicspath{{figures/}}

\begin{document}

\title{Characterizing Quantum-Dot Blinking Using Noise Power Spectra}

\author{Matthew Pelton}
 \email{pelton@uchicago.edu}

\author{David G. Grier}

\author{Philippe Guyot-Sionnest}

\affiliation{James Franck Institute, University of Chicago, 5640 S. Ellis Ave., Chicago, IL 60637}

\date{\today}

\begin{abstract}
Fluctuations in the fluorescence from macroscopic ensembles of colloidal 
semiconductor quantum dots have 
the spectral form of $1/f$ noise.  
The measured power spectral density reflects the known
fluorescence intermittency of individual dots
with power-law distributions of ``on'' and ``off''
times, and can thus serve as a simple method for characterizing such blinking behavior.

\end{abstract}

\pacs{78.67.Bf, 81.07.Ta, 05.40.-a}


\maketitle

Semiconductor nanocrystals, or quantum dots (QDs), are valued for their
unique optical properties.  They can exhibit bright, long-lived
fluorescence, with an emission wavelength that is simply tuned by
changing the size of the nanocrystal.  This property makes them
promising, for example, as biological labels, and as the active
medium in light-emitting diodes or lasers.  
However, such applications
may be compromised by large-amplitude fluctuations in the intensity
of the QD fluorescence.
Optical microscopy of immobilized QDs has
shown a noisy blinking behavior, with the dots
alternating between ``on'' (fluorescing) and ``off'' (non-fluorescing)
states \cite{Nirmal96}.  Although results from different groups show some
differences, many measurements under different conditions have shown
similar power-law distributions for on and off periods \cite{Kuno01,Shimizu01,Brokman03}.
For a particular nanocrystal system and a particular experiment,
the blinking statistics are robust, and are independent of the particular
nanocrystal being observed.  While previous studies have focussed on
single QDs, this Letter shows that complementary information on the statistics of fluorescence
fluctuations is provided by measurements on macroscopic ensembles
of QDs, allowing a new perspective on the phenomenon of QD blinking.

In particular, we measured the power spectral density of
fluctuations in the fluorescence from QD ensembles. 
Such noise spectra are commonly used
to obtain statistical information about a wide variety of systems.
Applying this technique to QDs shows
that the QD fluorescence fluctuations have the form of $1/f$ noise.
The ensemble $1/f$ noise can be understood as the incoherent
sum of the noise spectra of individual QDs, all of which exhibit identical
$1/f$-type noise.  The ensemble noise measurements thus provide a simple,
rapid technique to obtain information about the blinking statistics of
individual QDs.  It can be applied to a wide range of environments, 
including those not previously accessible to experiment.

Our ensemble noise measurements were made on monodisperse CdSe QDs, prepared
using established methods \cite{Murray93}, with a fluorescence maximum
around a wavelength of 610 nm.  We investigated both bare CdSe nanocrystals and nanocrystals
capped with ZnS shells \cite{Hines96}; our results were essentially
identical for both types.  The noise measurements were made
by exciting the QDs with a
diode-pumped, frequency-doubled Nd:YVO$_4$ laser, 
which has a stable output at a wavelength of 532 nm. The
fluorescence was collected perpendicular to the excitation beam using
an optical fiber bundle.  The emission was then sent through a
dichroic mirror, to eliminate scattered laser light, and was detected
with a silicon photodiode.  The photodiode output was sent to a
low-noise amplifier, whose
input incorporates a high-pass filter with a cut-on at 1 Hz; this
eliminates the dots' mean fluorescence, so that only small-signal
fluctuations are amplified.  The amplified output was sent to a
digital signal analyzer, which
calculated the associated power spectral density.  The digitally
sampled time series were analyzed in blocks of 1600 points; 250 blocks
were combined using RMS averaging to give the averaged time series $I(t)$.
The power spectral density was then calculated as 
$S(f) =\left| {\mathcal F} \left\{ i(t) \right\} \right| ^2$, where
${\mathcal F}$ represents the Fourier transform, and $i (t) = I(t)
- \langle I \rangle$ is the instantaneous deviation of the intensity
from the mean $\langle I \rangle$.

We recorded power spectra between frequencies of 200 Hz and 3 kHz.  At lower
frequencies, fluorescence fluctuations are overwhelmed by noise in the excitation
laser, while, at higher frequencies, they are overwhelmed by noise in the detection apparatus
(photodiode, amplifier, and signal analyzer).  
By varying the incident laser power, we verified that the measured power spectral
density is proportional to the excitation power.  This means that fluctuations
in the laser power at a particular frequency will result in fluctuations in
the fluorescence intensity at the same frequency.  In order to remove these
effects, which do not reflect the blinking dynamics of the QDs, we
divided the power spectrum of the dots by the measured power spectrum of the laser.

\begin{figure}[tbhp]
  \centering
\begin{center} \includegraphics[height=2.75in]{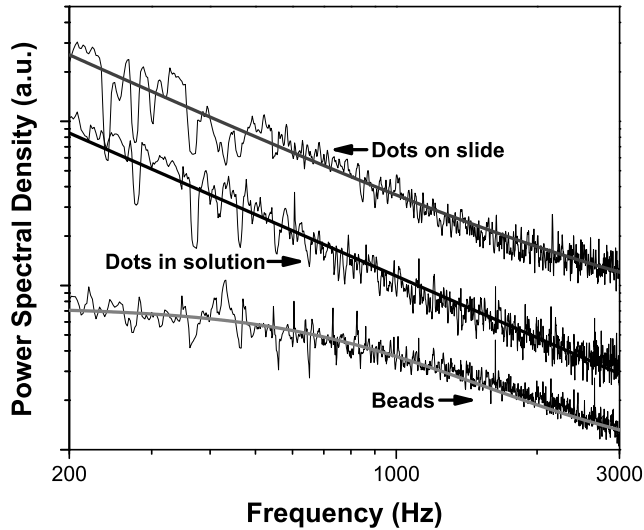} \end{center}
  \caption{Power spectral densities of fluctuations in the fluorescence from
  macroscopic ensembles of quantum dots, deposited on a glass slide
  and dissolved in chloroform, and from fluorescently dyed polystyrene
  beads.  The spectra have been offset by arbitrary scale factors for clarity.
  Thick lines are fits to the power spectra; the dot
  spectra are fit to power laws, while the dye spectrum is fit to a
  Lorentzian.} 
\label{dots-dye-laser}
\end{figure}

For the first measurements, we deposited QDs
in a dense layer on a glass slide, providing an environment similar to that
experienced by dots in microscopy experiments.
The measured noise spectrum is shown in
Fig.~\ref{dots-dye-laser}, and can be seen to have the form of $1/f$ noise over
the experimentally accessible bandwidth.
More specifically, it can be fitted using a relationship of the form
$S(f) = A f^{\nu - 2} + B$, where $f$ is the frequency, $A$ is an arbitrary
proportionality constant, $B$ is an additive offset representing the
instrumental noise floor, and $\nu$ is the fitted spectral exponent 
characterizing the noise-generating process.
In this case, $\nu = 0.70 \pm 0.02$.

$1/f$ noise spectra arise in a wide array of systems, 
including conductance fluctuations of metals and semiconductors,
magnetization of spin glasses, financial time series, and biological ion
channels.  They are qualitatively different from the spectrum obtained for
simple molecular systems, as we confirmed by measuring 
the power spectrum of an ensemble of
polystyrene microspheres doped with a green fluorescent dye (Duke
Scientific, part number G300).  This noise spectrum, shown in
Fig.~\ref{dots-dye-laser}, can be fit using a single Lorentzian 
$S(f) = A/(f^2 + f_0^2) + B$, with a width $f_0 = 1.8~\unit{kHz}$.
The Lorentzian noise spectrum can easily be understood as the
result of a fluctuation process with a single, characteristic
time scale; in this case, it may reflect shelving in a dark triplet state.
The difference between this spectrum and the QD spectrum verifies that the
measured $1/f$-type power spectrum is characteristic
of the QDs.

The ability to characterize the statistics of fluorescence fluctuations in
the ensemble means that immobilization of the QDs is not necessary.
We therefore made a second measurement, for which
we dissolved an ensemble of QDs in chloroform.
We illuminated a large volume 
(${\cal O}(1)~\unit{mm^3}$), so that the variation in observed particle number due
to diffusion was negligible.  The measured spectrum is shown in
Fig.~\ref{dots-dye-laser}, and can be seen to have the same form as the
spectrum of dots on glass, with an equivalent fitted exponent of $\nu = 0.75 \pm 0.03$.
The observed insensitivity of the blinking statistics to the QD environment
is surprising, in the light of currently proposed blinking mechanisms \cite{Verbeck02,Kuno03}.

Although no unified model of $1/f$ noise exists, it is
often attributed to the collective effect of a broad distribution
of independent processes with different characteristic times \cite{Weissman88}.
$1/f$ noise in QD fluorescence fluctuations, by contrast, appears to be 
an intrinsic property of the individual dots. 
This distinction can be made clear by considering how individual 
QD spectra contribute to our measurements.
By the Wiener-Khinchin theorem, the
power spectral density is equal to the Fourier transform of the intensity
autocorrelation function.  Since the total intensity emitted by an
ensemble of dots is the sum of the intensities $i_n(t)$ emitted by the
individual dots, the autocorrelation function of the total emission is
\begin{equation}
  G^{(2)}(\tau) = \sum_n \langle i_n(t) \, i_n(t+\tau) \rangle + 
  \sum_{m \neq n} \langle i_m(t) \, i_n(t+\tau) \rangle,
\end{equation}
where angle brackets indicate averages over time $t$.  The second
term describes cross-correlations of fluctuations from from different
QDs, and vanishes if they fluctuate independently.  In this case,
all that remains is the sum of the autocorrelation
functions for the individual QDs.  These autocorrelations are
equal to the Fourier transforms of the individual QD power spectra,
so the power spectrum of the total intensity is simply the sum of the
single-dot spectra.  If all dots fluctuate with the same statistics, then this spectrum 
also provides insights into the mechanism of single-dot fluorescence intermittency.

This was confirmed by monitoring the fluorescence from single QDs, and combining their emission
numerically in order to obtain the corresponding ensemble behavior.
A sparse layer of dots with emission maxima
around a wavelength of 535 nm was deposited on a glass microscope coverslip.
The dots were excited from the opposite side of the coverslip, through an
oil-immersion microscope objective, using light with a wavelength
of $480 \pm 40~\unit{nm}$.  Light emitted from the QDs was collected
through the same objective, separated from the excitation light
using a dichroic mirror, and isolated using a bandpass filter with
a center wavelength of 535 nm and a bandwidth of 30 nm.  The light
was imaged onto a cooled CCD camera, 
and the intensities from
particular QDs were monitored over approximately 13
minutes, with a time resolution of 25 ms.

\begin{figure}[tbhp]
  \centering
\begin{center} \includegraphics[height=2.75in]{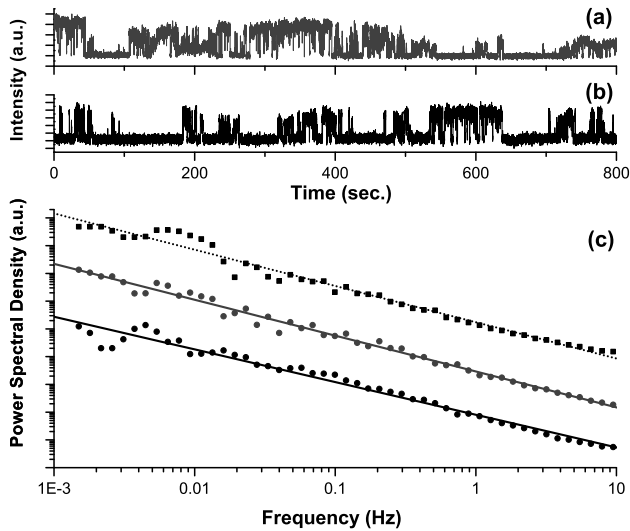} \end{center}
  \caption{(a),(b): Fluorescence intensity from two
  individual quantum dots as a function of time.  (c): Log-log
  plot of the power spectral densities calculated from the above time
  traces (circles), together with power-law fits (solid lines).  The
  two power spectra are offset by arbitrary scale factors for clarity.
  Also shown the is power spectral density calculated
  from the sum of 20 single-dot time series (squares).
  Calculated power spectral densities have been grouped in logarithmic
  frequency bins.} 
\label{dot}
\end{figure}

Time traces for two specific QDs are shown in Fig.~\ref{dot}(a) and (b).  Power
spectral densities were calculated from these time traces, after
subtracting the mean values and multiplying by the Hann window function 
$W(t) =\sin^2 (\pi t / T)$, where $T$ is the total duration of the data set
\cite{Press92}.  Resulting spectra are shown in
Fig.~\ref{dot}(c);  they can be seen to have the same $1/f$ type form as the
ensemble power spectra.  Fitting the two sample single-dot spectra to an inverse power
law over the entire bandwidth of the measurement yields
$\nu = 0.71$ and $\nu = 0.83$, respectively.
Comparable results were obtained for all 20 QDs studied, with a mean
value of $\nu = 0.74 \pm 0.09$.  This exponent is indeed the same as that obtained
for the QD ensembles.

Further confirmation that the ensemble measurement reflects single-dot statistics
is provided by adding together the measured time traces from the 
individual QDs point by point, and calculating the power spectrum from
the time trace of the sum (shown in
Fig.~\ref{dot}(c)).  The fitted exponent is $\nu = 0.69 \pm 0.03$, in
agreement with the average exponent for the individual QD spectra.
This is different from the case of
electronic $1/f$ noise, for instance, where the ensemble spectrum is an average of
individual fluctuators that each display very different dynamics.
In our case, the ensemble measurements can provide direct information
about the fluctuation statistics of individual QDs.

Ensemble power spectra thus provide a useful tool to complement
characterization techniques based on single-dot measurements.
In particular, our power spectra are consistent with previously reported distributions of on
and off periods for individual QDs.
The probability density functions of blinking periods $T_{\text{on/off}}$ 
have been observed to follow an inverse power
law $\Psi(T_{\text{on/off}}) \propto
(T_{\text{on/off}})^{-1-\nu}$ \cite{Kuno01}.  Power-law distributions of this
type, where $\nu$ lies between zero and one, can be converted into power spectra using the
mathematical tools devised to describe L\'{e}vy walks.  Specifically,
we impose lower and upper cutoff durations $T_{\text{min}}$ and
$T_{\text{max}}$, respectively, and assume that the probability
density falls abruptly to zero outside these limits.  Experimentally,
$T_{\text{min}}$ can be associated with the measurement time
resolution, and $T_{\text{max}}$ with the total duration of the
measurement.  We ignore any intensity fluctuations within the on and off
states, treating the intensity as if it switches discretely between
zero and a value $I_o$.  If we assume, for simplicity,
that the same exponent $\nu$ can be used to describe on
and off times, then the time series has a power
spectral density $S(f) \propto (1/\langle T \rangle) f^{\nu - 2}$,
where $\langle T \rangle$ is the mean on/off duration \cite{Geisel95}.
The cutoff times $T_{\text{min}}$ and $T_{\text{max}}$ enter only
through an additional proportionality constant, so the form of
the power spectrum is independent of the details of the measurement.
The value of $\nu$ that we have obtained from the power spectrum is consistent
with the exponents previously obtained by different researchers from
probability distributions of bright and dark times
\cite{Kuno01,Shimizu01,Brokman03}.  Since previous measurements of blinking-time
distributions have covered the bandwidths of both our measurements \cite{Kuno01},
we can assume that the power spectra we have measured for QD ensembles
reflects the same underlying statistics as the individual QD spectra.

Similar results would be obtained if the on and off durations followed different
probability distributions.  For example, if they followed power laws with
different exponents, the larger value of $\nu$ would dominate the power spectrum;
similarly, if one distribution followed an exponential distribution while the other followed
a power law, the power spectrum would still be a power law \cite{Verbeck03}.  
In other words, power spectral densities cannot separately characterize
on and off blinking statistics.  On the other hand, they
can be measured on ensembles, allowing easy characterization in
just a few minutes.  Even for measurements on single QDs, unambiguous
power spectra can be calculated using well-established methods \cite{Press92}.
By contrast, extracting the distribution of blinking periods from the time 
series is less straightforward.  For example, even while the QD is emitting light, its
fluorescence intensity can still vary substantially, making it
difficult to establish a clear threshold between the on and
off states.  As well, blinking events shorter than the time
resolution of the measurement will be missed, biasing the probability
density function towards longer times; this artifact cannot be
corrected without making an assumption about the distribution of
blinking periods below the time resolution.

Another complementary method of characterizing QD fluorescence dynamics
is to measure the autocorrelation of emitted photons;  this allows a wide
range of time scales to be covered \cite{Messin01}.
Fluorescence correlation spectroscopy (FCS), a particular type of 
autocorrelation technique, has been used
to study the fluorescence of QDs in solution \cite{Larson03}.  
In these measurements, a microscopic volume of a
solution containing QDs is illuminated, and the autocorrelation
function of the detected fluorescence is measured as dots diffuse
through the illuminated region.  The authors of Ref.~\cite{Larson03}
were able to fit their results with a model based only on
single-particle diffusion, with no systematic deviation that would
need to be explained by blinking.  Their results can be reconciled
with ours by considering how blinking affects the autocorrelation
function.  The L\'{e}vy-walk model yields
$G^{(2)}(\tau) = A - B\tau^{1-\nu}/{\langle T \rangle}$, 
where $A = 1 - (\nu /(1-\nu)) (T_{\text{min}}/\langle T \rangle)$, and 
$B = T_{\text{min}}^\nu/(1-\nu)$.  This expression describes a correlation
function that is weakly dependent on delay time $\tau$ for short
$\tau$, and abruptly drops to zero as $\tau$ approaches the total
measurement time $T_{\text{max}}$ \cite{Verbeck02,Messin01}.  
The effects of diffusion in FCS can be approximated
by multiplying the above autocorrelation function with one describing
diffusion \cite{Maiti97}, taking $T_{\text{max}}$ to be the mean
diffusion time through the illuminated volume.  Since blinking gives a
nearly flat correlation function for times shorter than the diffusion
time, the combination is nearly indistinguishable from diffusion
alone.  By contrast, the power spectral density has the advantage of clearly
distinguishing blinking from diffusion, and is insensitive to
measurement parameters such as the time resolution and the total
measurement time.
Other ensemble measurement techniques can also provide information on
fluorescence fluctuations.
For example, the reversible decay in the
fluorescence signal from an ensemble of QDs has been shown to have a
purely statistical origin \cite{Brokman03}.  It has also been proposed that the
variance in the number of photons emitted by an ensemble of QDs should diverge
over time \cite{Jung02}.  

We have found that it is particularly practical to study the dynamics of QD blinking
by measuring the power spectral density of fluorescence
fluctuations.  The results are 
free from the ambiguities inherent in measurements based on
blinking time distributions and autocorrelation
functions.  Representative power spectra can be measured on ensembles of
dots, contrary to the popular wisdom that blinking studies require
isolation of single emitters.  This means that blinking can be
observed in cases where microscopy is impractical, such as QDs
in solution. 
Using lower-noise components, it should be straightforward
to extend the measurement bandwidth beyond that obtained in this first experiment.
Since the ensemble power spectrum can be measured
very quickly, it will be possible to rapidly characterize the
blinking behavior of different samples in different environments, eventually
leading to a better understanding of and control over the blinking mechanism.
Finally, we have used the same measurement technique to observe
the fluorescence dynamics of fluorescent microspheres;  indeed, the method we
have introduced should be applicable to any fluorophore that
exhibits blinking within the observable bandwidth.

\begin{acknowledgments} We would like to thank C.~Wang for fabricating the nanocrystal quantum dots, and V.~Bindokas
for valuable assistance with microscopic measurements.  This work was principally supported by the MRSEC program of the NSF
through Grant Number DMR-0213745, with additional support from NSF
Grant Number DMR-0304906 and the donors of the Petroleum Research Fund
of the American Chemical Society.
\end{acknowledgments}

\bibliographystyle{apsrev}
\bibliography{noise}

\end{document}